\pdfoutput=1
\RequirePackage{ifpdf}
\ifpdf 
\documentclass[pdftex]{sigma}
\else
\documentclass{sigma}
\fi

\newcommand{\De}{\Delta}
\newcommand{\de}{\delta}

\newcommand{\N}{{\mathbb N}}
\newcommand{\C}{{\mathbb C}}
\newcommand{\R}{{\mathbb R}}

\newcommand{\bP}{{\mathbb P}}

\newcommand{\cE}{{\cal E}}
\newcommand{\cA}{{\cal A}}

\newcommand{\cF}{{\cal F}}
\newcommand{\cG}{{\cal G}}

\numberwithin{equation}{section}

\newtheorem{Theorem}{Theorem}[section]

 { \theoremstyle{definition}

\newtheorem{Remark}[Theorem]{Remark} }

\begin{document}
\allowdisplaybreaks

\newcommand{\arXivNumber}{1903.09738}

\renewcommand{\thefootnote}{}

\renewcommand{\PaperNumber}{063}

\FirstPageHeading

\ShortArticleName{The Elliptic Painlev\'e Lax Equation vs.\ van Diejen's 8-Coupling Elliptic Hamiltonian}

\ArticleName{The Elliptic Painlev\'e Lax Equation\\ vs.\ van Diejen's 8-Coupling Elliptic Hamiltonian\footnote{This paper is a~contribution to the Special Issue on Elliptic Integrable Systems, Special Functions and Quantum Field Theory. The full collection is available at \href{https://www.emis.de/journals/SIGMA/elliptic-integrable-systems.html}{https://www.emis.de/journals/SIGMA/elliptic-integrable-systems.html}}}

\Author{Masatoshi NOUMI~$^{\dag}$, Simon RUIJSENAARS~$^{\ddag}$ and Yasuhiko YAMADA~$^{\dag}$}

\AuthorNameForHeading{M.~Noumi, S.~Ruijsenaars and Y.~Yamada}

\Address{$^{\dag}$~Department of Mathematics, Kobe University, Rokko, Kobe 657-8501, Japan}
\Address{$^{\ddag}$~School of Mathematics, University of Leeds, Leeds LS2 9JT, UK}

\ArticleDates{Received April 20, 2020, in final form June 26, 2020; Published online July 08, 2020}

\Abstract{The 8-parameter elliptic Sakai difference Painlev\'e equation admits a Lax formulation. We show that a suitable specialization of the Lax equation gives rise to the time-independent Schr\"odinger equation for the $BC_1$ 8-parameter `relativistic' Calogero--Moser Hamiltonian due to van Diejen. This amounts to a generalization of previous results concerning the Painlev\'e--Calogero correspondence to the highest level in the two hierarchies. }

\Keywords{Painlev\'e--Calogero correspondence; elliptic difference Painlev\'e equation; Ruijse\-naars--van Diejen Hamiltonian}

\Classification{39A06; 33E05}

\renewcommand{\thefootnote}{\arabic{footnote}}
\setcounter{footnote}{0}

\section{Introduction}

There exists a well-known correspondence between the nonlinear 4-parameter Painlev\'e diffe\-ren\-tial equation P(VI) and the linear 4-parameter Heun equation. The former admits a Lax formulation, and a suitable specialization of one of the linear Lax equations yields the Heun equation. This correspondence has come to be known as the Painlev\'e--Calogero correspondence. Indeed, the Heun equation can be viewed as the time-independent Schr\"odinger equation associated with the elliptic $BC_1$ Calogero Hamiltonian (also known as the 1-particle specialization of the Inozemtsev integrable particle system). This correspondence has been elaborated on from several viewpoints in~\cite{Taka01,Take04, ZZ12}.\footnote{For the higher rank models, there is also a correspondence \cite{Taka01}, where the Painlev\'e side of the correspondence is a kind of multi-dimensional extension different from the Garnier systems (see~\cite{BCR18} for recent developments).}

A decade ago, one of us (S.R.) conjectured that there might exist a similar
connection between van Diejen's~\cite{vD94} 8-parameter analytic difference operator and a Lax formulation for Sakai's~\cite{S01} 8-parameter elliptic Painlev\'e difference equation.
A~Lax formulation of the latter equation was obtained in \cite{Ra11, Y09} and then simplified in
\cite{NTY13} (see~\cite{KNY17} for a review). In this article we prove the conjecture by invoking the Lax equation that occurs in equation~(4.4) in the recent paper~\cite{Y17}.

Previous results of a similar nature were recently obtained by Takemura~\cite{Take17}. They involve various specializations with fewer than~8 parameters, making use of Lax formulations of the pertinent difference Painlev\'e equations in~\cite{Y11} and Jimbo--Sakai~\cite{JS96}.

Our main result amounts to a generalized Painlev\'e--Calogero correspondence at the highest level of both the Painlev\'e hierarchy and the rank-1 (or `one-particle') Calogero--Moser hierarchy. A recent survey of the Painlev\'e equations can be found in~\cite{KNY17}, detailing a great many specializations and confluence limits. To date there exists no similar exhaustive published version of the corresponding degenerations of the van Diejen Hamiltonian.

In this article we use the latter in the somewhat different guise of the operator~$A_+(\gamma;x)$ defined by equations~(1.8) and (3.1)--(3.7) in~\cite{R15}. (We shall detail its definition shortly.) It differs from van Diejen's operator by multiplicative and additive constants, yielding a version that gives rise to various parameter symmetries. In particular, there exists a natural similarity transform~$\cA_+(\gamma;x)$ (defined by equations~(1.23) and (3.12) of~\cite{R15}) that exhibits manifest symmetry under permutations and even sign changes of the 8 coupling-type parameters $\gamma_0,\ldots,\gamma_7$. Under certain restrictions, this $W(D_8)$ symmetry is shown to extend to a $W(E_8)$ spectral invariance of $\cA_+(\gamma;x)$, reinterpreted as a self-adjoint operator on a suitable Hilbert space~\cite{R15}. This is clearly analogous to the $W\big(E_8^{(1)}\big)$ symmetry of Sakai's elliptic Painlev\'e equation. Even so, just as the correspondence we establish, to date this state of affairs is not understood at a more fundamental level.

We proceed to specify the notations used in~\cite{ R15} and~\cite{Y17}, and to set up a translation between them. To this end we begin by comparing the two starting points, which involve distinct conventions for the elliptic gamma function. The convention in~\cite{Y17} is now used in most of the literature that employs the elliptic gamma function, namely,
\begin{gather}\label{Gamell}
\Gamma_{p,q}(z)\equiv \prod_{k,l= 0}^{\infty}\frac{1-z^{-1}p^{k+1}q^{l+1}}{1-zp^{k}q^{l}}.
\end{gather}
This yields the $q$-difference equation
\begin{gather*}
\Gamma_{p,q}(qz)=[z]\Gamma_{p,q}(z),
\end{gather*}
with $[z]$ the key building block of \cite{Y17},
\begin{gather}\label{z}
[z]=\prod_{m=0}^{\infty}\big(1-zp^m\big)\big(1-z^{-1}p^{m+1}\big).
\end{gather}
Thus, $[z]$ can be viewed as a theta function, but it differs from the four Jacobi theta functions, inasmuch as it is neither even nor odd.

By contrast, in~\cite{ R15} the `original' definition of the elliptic gamma function~\cite{R97} is used, viz.,
\begin{gather}\label{Gell}
	G(r,a_+,a_-;x) \equiv \prod_{m,n=0}^\infty \frac{1-\exp \big({-}(2m+1)ra_+-(2n+1)ra_--2{\rm i}rx\big)}{1-\exp \big({-}(2m+1)ra_+-(2n+1)ra_-+2{\rm i}rx\big)}.
\end{gather}
Thus we need to substitute
\begin{gather}\label{pq}
p=\exp(-2ra_+),\qquad q=\exp(-2ra_-),
\end{gather}
and
\begin{gather}\label{zx}
z=\exp(2{\rm i}r(x+{\rm i}a)),\qquad a\equiv (a_++a_-)/2,
\end{gather}
on the r.h.s.\ of~(\ref{Gamell}) to arrive at the r.h.s.\ of~(\ref{Gell}). (The second convention has definite advantages for quantum-mechanical purposes, but its drawback is that it deviates from the Euler gamma function by not having its `first' simple pole at $x=0$, but at $x=-{\rm i}a$; by contrast, $\Gamma_{p,q}(z)$ has its `first' simple pole at $z=1$.)

For the purpose of this paper, the building block in this second setting is one of the two right-hand side functions arising from the analytic difference equations (henceforth called A$\De$Es)
\begin{gather}\label{Gades}
\frac{G(x+{\rm i}a_\delta/2)}{G(x-{\rm i}a_\delta/2)} = R_{-\delta}(x),\qquad \delta=+,-,
\end{gather}
namely,
\begin{gather}\label{Rdef}
R_+(x)=\prod_{m=0}^{\infty}[1-\exp(2{\rm i}rx-(2m+1) ra_+)][1-\exp(-2{\rm i}rx-(2m+1) ra_+)].
\end{gather}
Up to normalization, this is one of the Jacobi theta functions, cf.~\cite{WW73}. It is even, $\pi/r$-periodic, and has no real zeros; the only further property that needs to be often used in the sequel is the~A$\De$E
\begin{gather}\label{RDE}
R_+(x+{\rm i}a_+/2)=-\exp(-2{\rm i}rx)R_+(x-{\rm i}a_+/2),
\end{gather}
which can be deduced from~(\ref{Rdef}).

After the substitutions (\ref{pq}) and (\ref{zx}) on the r.h.s.\ of~(\ref{z}), we can compare to~(\ref{Rdef}) to obtain the key relation
\begin{gather}\label{zRrel}
[z] =R_+\left(\frac{1}{2{\rm i}r}\ln z-{\rm i}a_+/2\right)
\end{gather}
between the two building blocks.

We are now prepared to focus on the two equations at issue. Van Diejen's operator is of the form
\begin{gather}\label{A+}
A_+(\gamma;x)=V(\gamma;x)\exp(-{\rm i}a_-{\rm d}/{\rm d}x) + V(\gamma;-x)\exp({\rm i}a_-{\rm d}/{\rm d}x) +V_{\rm b}(\gamma;x).
\end{gather}
Here we have
\begin{gather}\label{V}
V(\gamma;x)\equiv \frac{\prod\limits_{\mu=0}^7R_{+}(x-{\rm i}\gamma_{\mu}-{\rm i}a_{-}/2)}{R_{+}(2x+{\rm i}a_{+}/2)R_{+}(2x+{\rm i}a_{+}/2-{\rm i}a_{-})},
\end{gather}
whereas $V_{\rm b}(\gamma;x)$ is an even elliptic function with periods $\pi/r$, ${\rm i}a_+$. It is uniquely determined up to a constant by specifying the residues at its (generically) simple poles, which occur at the zeros of the factors $R_+(\pm 2x+{\rm i}a_+/2-{\rm i}a_-)$ of the shift coefficients $V(\gamma;\pm x)$. (Thus we get eight poles in a period parallelogram.) We shall define the residues later on, cf.~\eqref{xn}--\eqref{rho}.

The operator $A_+(\gamma;x)$ is viewed as a quantum-mechanical one, in the sense that the pertinent equation to solve is the time-independent Schr\"odinger equation
\begin{gather}\label{Schr}
(A_+(\gamma;x)-E)\psi(x)=0.
\end{gather}
By contrast, at face value the equation in \cite{Y17} ($N=3$ case of $L_1$ in equation~(4.4)) has a different structure. Indeed, it is given in the form of a $q$-difference equation that is not of this Schr\"odinger type, namely,
\begin{gather}\label{Yeq}
W_-(z)y(z/q) + W_+(z)y(qz) -R(z)y(z)=0.
\end{gather}
Here we have
\begin{gather}\label{Wm}
W_-(z)\equiv A(k/z)B(z)\cF(qz)\big[k/q^2z^2\big],\\
\label{Wp} W_+(z)\equiv A(qz)B(k/qz)\cF(z)\big[k/z^2\big],
\end{gather}
with
\begin{gather}
A(z)\equiv \prod_{j=1}^4 [z/a_j],\qquad B(z)\equiv \prod_{j=1}^4 [z/b_j],\qquad
\label{defF}
\cF(z)\equiv Cz[z/\lambda][k/z\lambda].
\end{gather}
The parameter $C$ can be viewed as a gauge parameter and the parameter~$k$ arises from the extended affine~$E_8$ Weyl group picture associated with the elliptic Sakai equation. Also, $\lambda$~encodes one of the two initial values for this equation, (\ref{Yeq})~being one of the two corresponding Lax equations, cf.\ equation~(4.4) of~\cite{Y17}.

\begin{Remark}\label{remark1}
Sakai's Painlev\'e equation has a bi-rational form in algebraic coordinates $(f,g) \in \bP^1\times \bP^1$.
The coordinates $f$, $g$ and $\lambda$, $\xi_1$ (\ref{xiell}) are related as $f=f(\lambda)$, $g=g(\xi_1)$ where~$f(u)$,~$g(u)$ are certain elliptic functions. In terms of the coordinates $f$, $g$, the Lax equation~(\ref{Yeq}) has the geometric characterization as a bi-degree $(3,2)$ curve passing through certain 12 points (see \cite[Section~7]{KNY17} for example).
\end{Remark}

To unburden the exposition in this introductory section, we do not completely specify $R(z)$ for now. Indeed, its definition is somewhat involved, just like the definition of the additive potential $V_{\rm b}(\gamma;x)$. However, it is expedient to detail at this stage the building block of $R(z)$ that carries the dependence on the eight fixed points $a_j$, $b_j$, $j=1,\ldots,4$, viz.,
\begin{gather}\label{U}
U(z)\equiv A(z)B(z).
\end{gather}
In general terms, $R(z)$ also depends on the above function $\cF(z)$ and a further function $\cG(z)$ that involves an extra parameter $\xi_1$ encoding the second initial value for the elliptic Sakai equation. (The details of the definition of $R(z)$ can be found in equations~\eqref{RS}--\eqref{evol}.)

Now the problem at issue is whether we can recast the Lax equation (\ref{Yeq}) in an alternative form, such that upon a suitable specialization of $C$, $k$, $\lambda$ and $\xi_1$ it takes the form of the eigenvalue equation~(\ref{Schr}).

To begin with, it is clear from (\ref{zx}) and (\ref{pq}) that we can switch from $z$ to $x$ by setting
\begin{gather}\label{psiy}
\psi(x)=y(2{\rm i}r(x+{\rm i}a)),
\end{gather}
so that the multiplicative $q$-shifts in~(\ref{Yeq}) turn into the additive ${\rm i}a_-$-shifts in~(\ref{Schr}).

Next, we recall that the variable change $z\to x$ entails the relation~\eqref{zRrel} between the building block functions $[z]$ and $R_+(x)$. We shall often use~\eqref{zRrel} in the sequel. For example, we have occasion to invoke the formula
 \begin{align}
\big[k/qz^2\big]& =R_+\left(\frac{1}{2{\rm i}r}\big(\ln k -\ln q\big)-2(x+{\rm i}a)-{\rm i}a_+/2\right)\nonumber\\
& =R_+(2x-(\ln k)/2{\rm i}r+2{\rm i}a_-+3{\rm i}a_+/2),\label{kq}
\end{align}
where we used (\ref{pq}) and evenness. Likewise, we need the formulas
\begin{gather}\label{ab}
[z/a_j]=R_+(x-(\ln a_j)/2{\rm i}r +{\rm i}a_-/2),\qquad [z/b_j]=R_+(x-(\ln b_j)/2{\rm i}r +{\rm i}a_-/2),
\end{gather}
so as to connect the eight parameters $a_j$ and $b_j$ to the eight $\gamma_{\mu}$'s.

The resulting division of the eight $\gamma_{\mu}$'s into two groups of four is clearly arbitrary, since the function $V(\gamma;x)$~\eqref{V} is invariant under permutations of the $\gamma_{\mu}$'s. In fact, taking gauge transformations into account, we can also change the signs of the $\gamma_{\mu}$'s in $V(\gamma;x)$, in keeping with the above-mentioned $W(D_8)$ invariance of the similarity transform $\cA_+(\gamma;x)$ of $A_+(\gamma;x)$. Indeed, as we shall see in Section~\ref{section3}, the additive part $V_{\rm b}(\gamma;x)$ is manifestly invariant under permutations and an even number of sign changes.

This state of affairs is crucial for the correspondence we aim to establish. We shall opt for choosing
\begin{gather}\label{abgam}
a_j=q\exp(-2r\gamma_{j-1}),\qquad b_j=q\exp(-2r\gamma_{j+3}),\qquad j=1,2,3,4,
\end{gather}
so that we get from~\eqref{U} and~\eqref{ab} the permutation-invariant function
\begin{gather}\label{U2}
U(z)=\prod_{j=1}^4[z/a_j][z/b_j]= \prod_{\mu=0}^7 R_+(x-{\rm i}\gamma_{\mu}-{\rm i}a_-/2).
\end{gather}
When we now consider the factors $[z/b_j]$ in $W_-(z)$~\eqref{Wm} and $[qz/a_j]$ in $W_+(z)$~\eqref{Wp}, we see upon comparison to $V(\gamma;x)$ and $V(\gamma;-x)$ that we need to switch the sign of $\gamma_0,\ldots,\gamma_3$ in $V(\gamma;x)$ to obtain equality of these four factors.

We can readily change the sign of $\gamma_{\mu}$ in $V(\gamma;x)$ by using a gauge factor
\begin{gather*}
G_{\mu}(x)\equiv G(x+{\rm i}\gamma_{\mu})/G(x-{\rm i}\gamma_{\mu}).
\end{gather*}
Specifically, using the $G$-A$\De$E~(\ref{Gades}) with $\de=-$, we get
 \begin{gather*}
 \frac{1}{G_{\mu}(x)}\exp(-{\rm i}a_-{\rm d}/{\rm d}x)G_{\mu}(x)=\frac{R_+(x-{\rm i}\gamma_{\mu}-{\rm i}a_-/2)}{R_+(x+{\rm i}\gamma_{\mu}-{\rm i}a_-/2)}\exp(- {\rm i}a_-{\rm d}/{\rm d}x),
 \end{gather*}
and
 \begin{gather*}
 \frac{1}{G_{\mu}(x)}\exp( {\rm i}a_-{\rm d}/{\rm d}x)G_{\mu}(x)=\frac{R_+(x+{\rm i}\gamma_{\mu}+{\rm i}a_-/2)}{R_+(x-{\rm i}\gamma_{\mu}+{\rm i}a_-/2)}\exp({\rm i}a_-{\rm d}/{\rm d}x),
 \end{gather*}
from which our assertion easily follows. As a consequence, we can employ a transformation with $(G_0G_1G_2G_3)(x)$ so as to transform the shift coefficient
\begin{gather}\label{Vtil}
\tilde{V}(\gamma;x)\equiv \frac{\prod\limits_{\mu=0}^3R_{+}(x+{\rm i}\gamma_{\mu}-{\rm i}a_{-}/2)\prod\limits_{\mu=4}^7R_{+} (x-{\rm i}\gamma_{\mu}-{\rm i}a_{-}/2)}{R_{+}(2x+{\rm i}a_{+}/2)R_{+}(2x+{\rm i}a_{+}/2-{\rm i}a_{-})},
\end{gather}
 back to $V( \gamma;x)$~\eqref{V}.

Besides the option to perform gauge transformations so as to change the shift part, we can also multiply the equation~\eqref{Yeq} by a suitable function. In the following sections we shall exploit this freedom, alongside further reparametrizations, so as to relate it to the Schr\"odinger equation~\eqref{Schr}. We do this in two stages, which enables us to get a better picture of the cogency of the correspondence at hand. In Section~\ref{section2} we first focus on the shift part of the equations. This reveals that there is a very limited choice in specializing the shift part of~\eqref{Yeq} so that the gauge-transformed shift part of the van Diejen Hamiltonian~\eqref{A+} arises. Indeed, already in this first stage it is remarkable that this is feasible.

The acid test is then whether the pertinent choices are compatible with having the same additive part. This test is addressed and passed in Section~\ref{section3}. In Section~\ref{section4} we add further insights. In particular, we show that the somewhat implicit parametrization of the function~$R(z)$ in~\eqref{Yeq} can be made fully explicit.

\section{The relation between the shift parts}\label{section2}

As a first step, we should complete the connection between the factors of $W_{\pm}(z)$ and $\tilde{V} (\gamma;\pm x)$ \eqref{Vtil} that depend on the parameters $a_j$, $b_j$ and~$\gamma_{\mu}$.
Proceeding as before, we get by using~(\ref{abgam}) for the factors $[k/za_j]$ in $W_-(z)$~\eqref{Wm} and the factors
$[k/qzb_j]$ in $W_+(z)$~\eqref{Wp} the relations
\begin{gather*}
[k/za_j]=R_+(x+{\rm i}\gamma_{j-1}-(\ln k)/2{\rm i}r+3{\rm i}a_-/2+{\rm i}a_+),\qquad j=1,2,3,4,
\end{gather*}
and
\begin{gather*}
[k/qzb_j]=R_+(x+{\rm i}\gamma_{j+3}-(\ln k)/2{\rm i}r+5{\rm i}a_-/2+{\rm i}a_+),\qquad j=1,2,3,4.
\end{gather*}
Clearly, the simplest way to turn this into the factors $R_+(x+{\rm i}\gamma_{j-1}-{\rm i}a_-/2)$ in $\tilde{V}(\gamma;x)$ and the factors $R_+(x+{\rm i}\gamma_{j+3}+{\rm i}a_-/2)$ in $\tilde{V}(\gamma;-x)$ is to choose $k$ equal to $pq^2$. Indeed, this yields equality of the pertinent 4-parameter factors, hence completing the connection between the 8-parameter factors in $W_{\pm}(z)$ and $\tilde{V} (\gamma;\pm x)$.

At this point it should be stressed that we do not know any a priori reason for the parameter~$k$ to be equal under these two distinct determinations. Put differently, at this rather early stage we could already have hit an unsurmountable snag in setting up the desired correspondence.

Requiring from now on
\begin{gather*}
k=pq^2=\exp(-2ra_+)\exp( -4ra_-),
\end{gather*}
we get from (\ref{kq})
\begin{gather}\label{rep1}
\big[k/qz^2\big]=R_+(2x+{\rm i}a_+/2),
\end{gather}
and likewise
\begin{gather}\label{rep2}
\big[k/q^2z^2\big]= R_+(2x+{\rm i}a_+/2+{\rm i}a_-),\qquad \big[k/z^2\big]= R_+(2x+{\rm i}a_+/2-{\rm i}a_-),
\end{gather}
as is readily verified.

We proceed by dividing (\ref{Yeq}) by the product
\begin{gather}
\big[k/z^2\big]\big[k/qz^2\big]\big[k/q^2z^2\big]\nonumber\\
\qquad{} = R_+(2x+{\rm i}a_+/2-{\rm i}a_-)R_+(2x+{\rm i}a_+/2)R_+(2x+{\rm i}a_+/2+{\rm i}a_-).\label{prod1}
\end{gather}
 Accordingly, the shift part can be rewritten as (recall~(\ref{psiy}))
 \begin{gather}\label{shi1}
 \cF(qz)\tilde{V}(\gamma;x)\psi(x-{\rm i}a_-) +q\cF(z){\rm e}^{8{\rm i}rx}\tilde{V}(\gamma;-x)\psi(x+{\rm i}a_-),
 \end{gather}
 where we used the $R_+$-A$\De$E~(\ref{RDE}).

Next, we put $ \lambda =q\nu$,
so that (\ref{defF}) becomes
\begin{gather*}
\cF(z)=Cz[z/q\nu][pq/z\nu].
\end{gather*}
Then we divide (\ref{shi1}) by
\begin{gather}\label{prod2}
Cp^{-1}z^3 [z/q\nu][p/z\nu],
\end{gather}
so that we obtain the new shift part
 \begin{gather*}
 {\rm e}^{-4{\rm i}rx}\frac{[z/\nu]}{[z/q\nu]}\tilde{V}(\gamma;x)\psi(x-{\rm i}a_-) +{\rm e}^{4{\rm i}rx}\frac{[pq/z\nu]}{[p/z\nu]}\tilde{V}(\gamma;-x)\psi(x+{\rm i}a_-).
 \end{gather*}

As our final reparametrization, we now set $ \nu=\exp(-2r\gamma_8)$.
 Then we readily obtain
 \begin{gather*}
 \frac{[z/\nu]}{[z/q\nu]}=\frac{R_+(x-{\rm i}\gamma_8+{\rm i}a_-/2)}{R_+(x-{\rm i}\gamma_8-{\rm i}a_-/2)},
 \end{gather*}
 and
 \begin{gather*}
 \frac{[pq/z\nu]}{[p/z\nu]}=\frac{R_+(x+{\rm i}\gamma_8-{\rm i}a_-/2)}{R_+(x+{\rm i}\gamma_8+{\rm i}a_-/2)}.
 \end{gather*}

 As a result, we need only specialize $\gamma_8$ to one of $\gamma_{\mu}$, $\mu=4,5,6,7$, in $\tilde{V}(\gamma;\pm x)$ to arrive at coefficients of the same structure, the only difference being that the pertinent $\gamma_{\mu}$ is replaced by $\gamma_{\mu}-a_-$. When we ignore the plane wave factors, we therefore get the shift part of a (gauge-transformed) van Diejen operator with
 \begin{gather}\label{gamshift}
 \gamma_{\mu}\to\gamma_{\mu} -a_-.
 \end{gather}

 To be sure, we should still take care of the plane waves, but they can be removed by a second gauge transformation. The simplest way to do so is by using a gauge function of the form $\exp \big(c_1x^2+c_2x\big)$. However, this function is not $\pi/r$-periodic, which seems `unnatural'.

 In order to avoid this problem, we shall use the $\pi/r$-periodic gauge function
 \begin{gather}\label{gau1}
 g(x)\equiv R_-(x-{\rm i}a_-/2)R_-(x+{\rm i}a_-/2).
 \end{gather}
 (Here $R_-$ is given by the r.h.s.\ of~\eqref{Rdef} with $a_+\to a_-$.) Indeed, we have
 \begin{gather}\label{gau2}
 \frac{1}{g(x)}\exp(\mp {\rm i}a_-{\rm d}/{\rm d}x)g(x)=q^{-1}{\rm e}^{\pm 4{\rm i}rx}\exp(\mp {\rm i}a_-{\rm d}/{\rm d}x).
 \end{gather}
 Thus it remains to multiply the equation~\eqref{Yeq} by $q$ to obtain what we want.

Admittedly, the function $g(x)$ also seems somewhat `alien' in the Lax context. But at least it is quite natural in the van Diejen context, in view of the modular symmetry of the latter~\cite{R15}. In any event, we see no way to avoid this gauge transformation in establishing the desired connection for the shift parts.

\section{The relation between the additive parts}\label{section3}

Let us now take stock of our findings in the previous section. We should divide the equation~(\ref{Yeq}) by the product
\begin{gather}\label{P}
P(z)\equiv Cp^{-1}q^{-1}z^3[z/\lambda] [k/qz\lambda]\big[k/z^2\big]\big[k/qz^2\big]\big[k/q^2z^2\big],
\end{gather}
cf.~(\ref{prod2}) and (\ref{prod1}), the factor $q^{-1}$ resulting from the gauge transformation~(\ref{gau1})--(\ref{gau2}). In the resulting equation, we should trade the parameters $p$ and $q$ for $a_+$ and $a_-$ by using~(\ref{pq}), the variable $z$ for $x$ by using~(\ref{zx}), and we should reparametrize $a_j$, $b_j$ via (\ref{abgam}). Finally, we need to choose
\begin{gather}\label{klam}
k=pq^2,\qquad \lambda =q\exp(-2r\gamma_8).
\end{gather}

The acid test is now whether the ratio $-R(z)/P(z)$ becomes equal to $V_{\rm b}(\gamma;x)-E$ under the above reparametrizations, once we choose (say)\footnote{In terms of the variables $(f,g)$ in Remark~\ref{remark1}, this specialization of the variable~$\lambda$ corresponds to putting \mbox{$f=f(b_4)$}. Under this specialization, the equation~(\ref{Yeq}) is factored by $g-g(b_4)$ due to the geometric characterization. As we will see below, the quotient gives the van Diejen operator with $E=(a g+b)/(g-g(k/b_4))$ ($a$,~$b$~are constants).}
\begin{gather}\label{87}
\gamma_8=\gamma_7,
\end{gather}
and substitute
\begin{gather*}
\gamma_7\to\gamma_7 -a_-,
\end{gather*}
in $V_{\rm b}(\gamma;x)$, cf.~(\ref{gamshift}). The expectation is that the additional parameters occurring in $-R(z)/P(z)$ should give rise to the eigenvalue $E$, as a counterpart of the second P(VI) initial value becoming proportional to the eigenvalue in the Heun equation in elliptic Schr\"odinger form under the Painlev\'e--Calogero correspondence.

Before discussing $R(z)$ as given in~\cite{Y17}, let us define $V_{\rm b}(\gamma;x)$. We shall not follow~\cite{R15} in doing so, but rather use the characterization that can be found in~\cite{RR13} (see also \cite{KNS09} for an explicit form of the operators together with their relation to original one). This is a definition that specifies $V_{\rm b}(\gamma;x)$ uniquely up to an arbitrary additive constant.

Specifically, $V_{\rm b}(\gamma;x)$ is an even elliptic function with periods $\pi/r$ and ${\rm i}a_+$, having eight (generically) simple poles in a period parallelogram. Modulo the elliptic lattice, these are located at $x=\pm x_n$, $n=0,1,2,3$, where
\begin{gather}\label{xn}
\{ x_0, x_1,x_2,x_3\} := -{\rm i}a_-/2 +\{0,\pi/2r,{\rm i}a_+/2,{\rm i}a_+/2+\pi/2r\}.
\end{gather}
The residues at these poles are given by
\begin{gather}\label{r0}
\rho_0\equiv \operatorname{Res} (x_0)=\eta \prod_{\mu}R_+({\rm i}\gamma_{\mu}),\\
\label{r1}
\rho_1\equiv \operatorname{Res} (x_1)=\eta \prod_{\mu}R_+({\rm i}\gamma_{\mu}+\pi/2r),\\
\label{r2}
\rho_2\equiv \operatorname{Res} (x_2)=\eta \exp\bigg({-}2ra_+-r\sum_{\mu}\gamma_{\mu}\bigg)\prod_{\mu}R_+({\rm i}\gamma_{\mu}+{\rm i}a_+/2),\\
\label{r3}
\rho_3\equiv \operatorname{Res} (x_3)=\eta \exp\bigg({-}2ra_+-r\sum_{\mu}\gamma_{\mu}\bigg)\prod_{\mu}R_+({\rm i}\gamma_{\mu}+{\rm i}a_+/2+\pi/2r),
\end{gather}
where
\begin{gather}\label{eta}
\eta \equiv \rho/2R_+({\rm i}a_-+{\rm i}a_+/2),
\end{gather}
with
\begin{gather}\label{rho}
\rho \equiv \lim_{x\to 0}\frac{x}{R_+(x+{\rm i}a_+/2)} = 1/2{\rm i}r\prod_{k=1}^{\infty} (1-\exp(-2kra_+))^2,
\end{gather}
cf.~\eqref{Rdef}. By evenness, the residues at $x=- x_n$ are then given by $-\rho_n$, $n=0,1,2,3$. Note that all of these residues are invariant under permutations and an even number of sign changes of $\gamma_0,\ldots,\gamma_7$.

\looseness=-1 In accordance with \cite[Section~4.3]{RR13}, the residues (\ref{r0}) and (\ref{r1}) are equal to minus the residues of
\begin{gather*}
V(\gamma;-x)= \frac{\prod\limits_{\mu=0}^7R_{+}(x+{\rm i}\gamma_{\mu}+{\rm i}a_{-}/2)}{R_{+}(2x-{\rm i}a_{+}/2)R_{+}(2x-{\rm i}a_{+}/2+{\rm i}a_{-})},
\end{gather*}
 (cf.~(\ref{V})) at $x_0$ and $x_1$, whereas the
 residues (\ref{r2}) and (\ref{r3}) should be multiplied by a factor{\samepage
\begin{gather}\label{mult_factor}
-\exp\bigg(2ra_-+2ra_++r\sum_{\mu}\gamma_{\mu}\bigg),
\end{gather}
to obtain the residues of $V(\gamma;-x)$ at $x_2$ and~$x_3$.\footnote{A similar relation between the residues of shift/additive terms is also known in the Lax operator side. However, the corresponding statement in~\cite[(iii) in Section~4]{Y17} is not literally correct, because of the disagreement of the quasi-periodicity of the shift/additive terms. This problem can be restored by a certain gauge transformation, or one should take into account multiplicative factors like~(\ref{mult_factor}).}}

As announced, we have now specified $V_{\rm b}(\gamma;x)$ up to an arbitrary additive constant. Turning to $R(z)$, it can be written
\begin{gather}\label{RS}
R(z)=\sum_{n=1}^3 S_n(z),
\end{gather}
where the summands are given by
\begin{gather}\label{S1}
S_1(z)\equiv U(z)\cF(qz)\cG(k/z)\big[k/q^2z^2\big]/\cG(z),
\\
\label{S2}
S_2(z)\equiv U(k/qz)\cF(z)\cG(qz)\big[k/z^2\big]/\cG(k/qz),
\\
\label{S3}
S_3(z)\equiv -\cF(z)\cF(qz)\overline{\cF}(z)\big[k/z^2\big]\big[k/qz^2\big]\big[k/q^2z^2\big]/\cG(z)\cG(k/qz),
\end{gather}
with $U(z)$ given by~\eqref{U} and $\cG(z)$ by
\begin{gather*}
\cG(z)\equiv z[z/\xi_1][z/\xi_2].
\end{gather*}
Furthermore, we have
\begin{gather}\label{xiell}
\xi_1\xi_2=\ell,
\end{gather}
where
\begin{gather}\label{kell}
k^2\ell^2=q\prod_{j=1}^4a_jb_j.
\end{gather}
(Just as $k$, the parameter $\ell$ stems from the extended affine Weyl group of $E_8$, cf.~\cite{KNY17}.)

It remains to define the function $\overline{\cF}(z)$. This is of the previous form~(\ref{defF}), but with `evolved' parameters $\overline{C},\overline{\lambda}$, and $\overline{k}$. The latter is simply given by
\begin{gather}\label{kev}
\overline{k}=k/q,
\end{gather}
so we have
\begin{gather*}
\overline{\cF}(z)=\overline{C}z\big[z/\overline{\lambda}\big]\big[k/qz\overline{\lambda}\big].
\end{gather*}
The evolution of $C$ and $\lambda$ is fixed by requiring
\begin{gather}\label{evol}
\cF(\xi_j)\overline{\cF}(\xi_j)\big[k/\xi_j^2\big]\big[k/q\xi_j^2\big]=\cG(k/\xi_j)\cG(k/q\xi_j)U(\xi_j),\qquad j=1,2,
\end{gather}
cf.\ \cite[Proposition~4.1]{Y17}. In view of (\ref{S1}) and~(\ref{S3}), this amounts to requiring that $R(z)$ have no poles at the zeros $z=\xi_1,\xi_2$, of~$\cG(z)$.

The crux of the simple $k$-evolution~(\ref{kev}) is that it guarantees that the function $R(z)$ is in fact holomorphic in the finite plane $\C^{*}$. This implies that
\begin{gather}\label{W}
W(x)\equiv R(\exp(2{\rm i}r(x+{\rm i}a)))
\end{gather}
is an entire $\pi/r$-periodic function, a feature that we shall return to in Section~\ref{section4}.

We proceed to calculate the summands
\begin{gather*}
T_n(x)\equiv S_n(\exp(2{\rm i}r(x+{\rm i}a))),\qquad n=1,2,3,
\end{gather*}
of $W(x)$ in terms of $R_+(x)$, using the above reparametrization~(\ref{klam}), and also setting
\begin{gather*}
\xi_i=q\exp(-2r\phi_i),\qquad i=1,2,
\qquad
\overline{\lambda}=q\exp(-2r\overline{\gamma_8}).
\end{gather*}

We have already calculated the factors containing the 8 parameters in the terms $T_n(x)$. Specifically, the factor coming from $U(z)$ is given by~\eqref{U2}, and from this we also obtain
\begin{gather*}
U(pq/z)= \prod_{\mu=0}^7 R_+(x+{\rm i}\gamma_{\mu}+{\rm i}a_-/2).
\end{gather*}
 To continue, we determine the remaining factors, using from now on the notation
 \begin{gather*}
 f(x\pm y):=f(x+y)f(x-y),
 \end{gather*}
 to shorten the resulting formulas
\begin{gather}
\cF(z)= C\exp(2{\rm i}r(x+{\rm i}a))R_+(x\pm {\rm i}\gamma_8-{\rm i}a_-/2) ,\nonumber\\
\cF(qz)= C\exp(2{\rm i}r(x+{\rm i}a))\exp(-2ra_-)R_+(x\pm {\rm i}\gamma_8+{\rm i}a_-/2) ,\nonumber\\
\label{Fbar2}
\overline{\cF}(z)= \overline{C}\exp(2{\rm i}r(x+{\rm i}a))R_+(x\pm ({\rm i}\overline{\gamma_8}+{\rm i}a_-/2)) ,
\\
\cG(z)= \exp(2{\rm i}r(x+{\rm i}a))\prod_{j=1}^2R_+(x-{\rm i}\phi_j-{\rm i}a_-/2),\nonumber\\
\cG(qz)= \exp(2{\rm i}r(x+{\rm i}a))\exp(-2ra_-)\prod_{j=1}^2R_+(x-{\rm i}\phi_j+{\rm i}a_-/2),\nonumber\\
\cG(pq/z)= \exp(-2{\rm i}r(x+{\rm i}a))\exp(-2ra_+-2ra_-)\prod_{j=1}^2R_+(x+{\rm i}\phi_j+{\rm i}a_-/2),\nonumber\\
\cG\big(pq^2/z\big)= \exp(-2{\rm i}r(x+{\rm i}a))\exp(-2ra_+-4ra_-)\prod_{j=1}^2R_+(x+{\rm i}\phi_j-{\rm i}a_-/2).\nonumber
\end{gather}
Using also (\ref{rep1}) and (\ref{rep2}), we finally get (recall $R_+(x)$ is even)
\begin{gather*}
T_1(x) = C{\rm e}^{-2{\rm i}rx}{\rm e}^{-ra_+}{\rm e}^{-5ra_-}R_+(x\pm {\rm i}\gamma_8+{\rm i}a_-/2)R_+(2x+{\rm i}a_+/2+{\rm i}a_-)\\
\hphantom{T_1(x) =}{} \times \prod_{\mu=0}^7R_+(x-{\rm i}\gamma_{\mu}-{\rm i}a_-/2) \prod_{j=1}^2\frac{R_+(x+{\rm i}\phi_j-{\rm i}a_-/2)}{R_+(x-{\rm i}\phi_j-{\rm i}a_-/2)},
\\
T_2(x)=T_1(-x),
\\
T_3(x)=-C^2\overline{C}{\rm e}^{6{\rm i}rx}{\rm e}^{-ra_+}{\rm e}^{-3ra_-}R_+(x\pm {\rm i}\gamma_8\pm {\rm i}a_-/2) R_+(x\pm ({\rm i}\overline{\gamma_8}-{\rm i}a_-/2))\\
\hphantom{T_3(x)=}{} \times R_+(2x+{\rm i}a_+/2)R_+(2x+{\rm i}a_+/2 \pm {\rm i}a_-)/\prod_{j=1}^2 R_+(x\pm ({\rm i}\phi_j+{\rm i}a_-/2)) .
\end{gather*}

Next, we calculate the denominator function (cf.~\eqref{P}),
\begin{gather*}
D(x)=P(\exp(2{\rm i}r(x+{\rm i}a))),
\end{gather*}
 by using the above factors
\begin{gather}
D(x)=C{\rm e}^{6{\rm i}rx}{\rm e}^{-ra_+}{\rm e}^{-ra_-} R_+(x\pm ({\rm i}\gamma_8+{\rm i}a_-/2))\nonumber\\
\hphantom{D(x)=}{} \times R_+(2x+{\rm i}a_+/2)R_+(2x+{\rm i}a_+/2\pm {\rm i}a_-).\label{D}
\end{gather}
Hence we obtain
\begin{gather}\label{Z}
-W(x)/D(x)=E(x)+E(-x)+V_{\rm e}(x)=:Z(x),
\end{gather}
where
\begin{gather}
E(x)\equiv - {\rm e}^{-8{\rm i}rx}{\rm e}^{-4ra_-}\frac{R_+(x-{\rm i}\gamma_8+{\rm i}a_-/2)}{R_+(x-{\rm i}\gamma_8-{\rm i}a_-/2)} \frac{\prod\limits_{\mu=0}^7R_+(x-{\rm i}\gamma_{\mu}-{\rm i}a_-/2)}{R_+(2x+{\rm i}a_+/2)R_+(2x+{\rm i}a_+/2-{\rm i}a_-)}\nonumber\\
\hphantom{E(x)\equiv}{} \times \prod_{j=1}^2\frac{R_+(x+{\rm i}\phi_j-{\rm i}a_-/2)}{R_+(x-{\rm i}\phi_j-{\rm i}a_-/2)},\label{E}
\\
\label{Ve}
V_{\rm e}(x)\equiv C\overline{C}{\rm e}^{-2ra_-}\frac{ R_+(x\pm {\rm i}(\overline{\gamma_8}-a_-/2))R_+(x\pm {\rm i}(\gamma_8-a_-/2))}{\prod_{j=1}^2 R_+(x\pm {\rm i}(\phi_j+a_-/2))}.
\end{gather}

We proceed to discuss these functions. The `extra' summand $V_{\rm e}(x)$ of the additive part $Z(x)$ is manifestly an even elliptic function. Moreover, we have
\begin{gather}
\frac{E(x+{\rm i}a_+/2)}{E(x-{\rm i}a_+/2)}= \exp(8ra_+)\exp(2ra_-)\frac{\prod\limits_{\mu=0}^7\exp(-2r\gamma_{\mu}-ra_-)}{\exp(4ra_+)\exp(-4ra_-)} \prod_{j=1}^2\exp(4r\phi_j)\nonumber\\
\hphantom{\frac{E(x+{\rm i}a_+/2)}{E(x-{\rm i}a_+/2)}}{}
=\exp(4ra_+)\exp(-2ra_-)\prod_{\mu=0}^7\exp(-2r\gamma_{\mu})\prod_{j=1}^2\exp(4r\phi_j).\label{Eade}
\end{gather}
Now with our reparametrizations, \eqref{kell} yields
\begin{gather*}
\ell^2=\exp(4ra_+)\exp(-10ra_-)\prod_{\mu=0}^7 \exp(-2r\gamma_{\mu}),
\end{gather*}
and then~\eqref{xiell} gives
\begin{gather}\label{phis}
\exp(-2r(\phi_1+\phi_2))=\exp(2ra_+)\exp(-ra_-)\prod_{\mu=0}^7 \exp(-r\gamma_{\mu}).
\end{gather}
Using this in \eqref{Eade}, we deduce that $E(x)$ is also elliptic.

Next, we study the remaining poles in the summands of $Z(x)$. By evenness, the poles coming from the denominator factors $R_+(2x+ {\rm i}a_+/2)$ and $R_+(2x- {\rm i}a_+/2)$ cancel. The remaining factors, however, do give rise to poles. In particular, for generic $\gamma_8$ we get poles depending on~$\gamma_8$.

We proceed to consider the special $\gamma_8$-choice~\eqref{87}. We already know from Section~\ref{section2} that for that choice the shift part amounts to that of a gauge-transformed van Diejen operator $A_+(\tilde{\gamma};x)$, with~$\tilde{\gamma}$ defined by
\begin{gather*}
\tilde{\gamma}_{\mu}\equiv \gamma_{\mu},\qquad \mu=0,\ldots,6,\qquad \tilde{\gamma}_7\equiv \gamma_7-a_-.
\end{gather*}
 Hence we should next compare the pole residues of $Z(x)$ and those of the additive part $V_{\rm b}(\tilde{\gamma};x)$ of $A_+(\tilde{\gamma};x)$. Also, by ellipticity and evenness we need only study the residues at $x=-x_n$, $n=0,1,2,3$.

For $x=-x_0={\rm i}a_-/2$, the summand $E(x)$~\eqref{E} with $\gamma_8=\gamma_7$ has a residue
\begin{gather*}
-\eta R_+({\rm i}\gamma_7-{\rm i}a_-)\prod_{\mu=0}^6R_+({\rm i}\gamma_{\mu}),
\end{gather*}
cf.~\eqref{eta}. Thus it equals the residue at $x=-x_0$ of $V_{\rm b}(\tilde{\gamma};x)$, cf.~\eqref{r0}. Likewise, using the $\pi/r$-periodicity of~$R_+(x)$, we see that the residues at $x=-x_1$ coincide.

 Let us next obtain the $E(x)$-residue for $x=-x_2={\rm i}a_-/2-{\rm i}a_+/2$. In this case~\eqref{E} yields the residue
 \begin{gather*}
 -{\rm e}^{-4ra_+} \rho \big[2{\rm e}^{-2ra_-}R_+({\rm i}a_-+{\rm i}a_+/2)\big]^{-1}R_+(-{\rm i}a_+/2 -{\rm i}\gamma_7+{\rm i}a_-)\prod_{\mu=0}^6R_+(-{\rm i}a_+/2-{\rm i}\gamma_{\mu})
 \\
 \qquad{} \times \prod_{j=1}^2\frac{R_+(-{\rm i}a_+/2+{\rm i}\phi_j)}{R_+(-{\rm i}a_+/2-{\rm i}\phi_j)},
 \end{gather*}
where we have used the $R_+$-A$\De$E~\eqref{RDE} and~\eqref{rho}. Using~\eqref{RDE} once more, this can be rewritten as
\begin{gather*}
-{\rm e}^{-4ra_+} {\rm e}^{2ra_-}\eta R_+({\rm i}\gamma_7-{\rm i}a_-+{\rm i}a_+/2)\prod_{\mu=0}^6R_+(i\gamma_{\mu}+{\rm i}a_+/2)
\cdot \exp(-2r(\phi_1+\phi_2)).
\end{gather*}
When we now use~\eqref{phis}, this becomes
\begin{gather*}
-{\rm e}^{-2ra_+}{\rm e}^{ra_-}\eta R_+({\rm i}\gamma_7-{\rm i}a_-+{\rm i}a_+/2)\prod_{\mu=0}^6R_+({\rm i}\gamma_{\mu}+{\rm i}a_+/2) \prod_{\mu=0}^7\exp(-r\gamma_{\mu}).
\end{gather*}
 Comparing this to~\eqref{r2}, we see that it equals the residue at $x=-x_2$ of $V_{\rm b}(\tilde{\gamma};x)$.

Basically the same calculation shows that the residues of $E(x)$ and $V_{\rm b}(\tilde{\gamma};x)$ at $x=-x_3$ are also equal. Therefore, the function
\begin{gather*}
V_{\rm b}(\tilde{\gamma};x)-Z(x)
\end{gather*}
has no poles, so by ellipticity it equals a constant, which we shall denote by~$E$. This constant~$E$, however, has a nontrivial dependence on the variable~$\phi_1$, which can be freely chosen (after which~$\phi_2$ is determined by~\eqref{phis}). It can be expected that when $\phi_1$ ranges over~$\C$, the corresponding $E$-values range over $\C$ as well. We shall study this further in~Section~\ref{section4}.

The upshot is that we have furnished the details of the correspondence between van Diejen's operator $A_+(\tilde{\gamma};x)$ and the Lax formulation for Sakai's elliptic Painlev\'e equation.\footnote{The generalization of the correspondence to the elliptic difference $n$-particle systems is an important future problem.} In the next section we add some further details and insights on this correspondence.

\section{Some supplements}\label{section4}

To begin with, let us confirm that $Z(x)$~\eqref{Z} has no poles for $\pm x$ equal to ${\rm i}\phi_1+{\rm i}a_-/2 -{\rm i}a_+/2$ and ${\rm i}\phi_2+{\rm i}a_-/2 -{\rm i}a_+/2$ modulo the elliptic lattice~$\Lambda$. To this end we recall that the quantities~$\overline{C}$ and $\overline{\gamma_8}$ are defined such that the function~$\overline{\cF}(z)$~\eqref{Fbar2} ensures the absence of poles of $R(z)$ at $z=\xi_n= q\exp(-2r\phi_n)$, $n=1,2$, cf.~\eqref{evol}. Now $P(z)$ has no zeros at these two $z$-values (generically), so $R(z)/P(z)$ has no poles there either. Recalling $z=\exp(2{\rm i}r(x+{\rm i}a))$, we see that the evolution requirement amounts to insisting that poles for
\begin{gather*}
x={\rm i}a_-/2 -{\rm i}a_+/2+{\rm i}\phi_n,\qquad n=1,2,
\end{gather*}
be absent in $Z(x)$. By evenness and ellipticity, it is then clear that $Z(x)$ indeed has no poles for $\pm x$ equal to ${\rm i}\phi_n+{\rm i}a_-/2 -{\rm i}a_+/2$ modulo~$\Lambda$. As a corollary, it follows from~\eqref{D} and~\eqref{Z} that $W(x)$~\eqref{W} is indeed an entire $\pi/r$-periodic function.

We proceed to study the evolution requirement in more detail. From~\eqref{E} and~\eqref{Ve} we see that for our variables and parameters it amounts to requiring
\begin{gather}\label{evreq}
\overline{C}R_+({\rm i}\phi_n +{\rm i}a_-/2 -{\rm i}a_+/2 \pm {\rm i}( \overline{\gamma_8}-a_-/2)) =c_n,
\end{gather}
where
\begin{gather}
 c_n\equiv C^{-1}\exp(8r\phi_n-4ra_+ +2ra_-) R_+({\rm i}\phi_1+{\rm i}\phi_2-{\rm i}a_+/2+{\rm i}a_-/2 \pm {\rm i}a_-/2)\nonumber \\
\hphantom{c_n\equiv}{} \times \frac{ \prod\limits_{\mu=0}^7R_+({\rm i}\phi_n-{\rm i}\gamma_{\mu}-{\rm i}a_+/2)}{R_+({\rm i}\phi_n\pm {\rm i}\gamma_8-{\rm i}a_+/2) },\qquad n=1,2.\label{cn}
 \end{gather}

We can solve the requirement~\eqref{evreq} (more or less) explicitly by making use of the family of even elliptic functions
\begin{gather}\label{Ed}
E(d;x)\equiv \frac{R_+(x\pm {\rm i}(d+a_+/2))}{R_+(x\pm {\rm i}a_+/2)},
\end{gather}
as follows. First, we note that we can rewrite~\eqref{evreq} as
\begin{gather}\label{ev2}
\overline{C}E(\psi;{\rm i}\alpha_n)=\frac{c_n}{R_+({\rm i}\alpha_n\pm {\rm i}a_+/2)}, \qquad n=1,2,
\end{gather}
where
\begin{gather}\label{an}
\alpha_n\equiv \phi_n +a_-/2 -a_+/2,\qquad n=1,2,
\end{gather}
and
\begin{gather*}
\psi \equiv \overline{\gamma_8}-a_-/2 -a_+/2.
\end{gather*}

Consider now the function
\begin{gather}\label{cE}
\cE(x)\equiv \frac{E(\alpha_2;x)}{R_+({\rm i}\alpha_1\pm {\rm i}(\alpha_2+a_+/2))}c_1 +
\frac{E(\alpha_1;x)}{R_+({\rm i}\alpha_2\pm {\rm i}(\alpha_1+a_+/2))}c_2.
\end{gather}
By construction, it satisfies
\begin{gather*}
\cE({\rm i}\alpha_n)=\frac{c_n}{R_+({\rm i}\alpha_n\pm {\rm i}a_+/2)}, \qquad n=1,2.
\end{gather*}
Now $\cE(x)$ is an even elliptic function with only two poles (at $x=0$) in a period parallelogram, so it must be of the form
\begin{gather*}
\cE(x)=\overline{C}\frac{R_+(x\pm {\rm i}(\psi +a_+/2))}{R_+(x\pm {\rm i}a_+/2)},
\end{gather*}
which determines the quantities $\overline{C}$ and $\psi$.
As a consequence, the quantities $\overline{C}$ and
\begin{gather*}
\overline{\gamma_8}= \psi+a_-/2+a_+/2,
\end{gather*}
yield a solution to~\eqref{evreq}, and we may substitute
\begin{gather}\label{subst}
\overline{C}R_+(x\pm {\rm i}(\overline{\gamma_8}-a_-/2))\to \cE(x)R_+(x\pm {\rm i}a_+/2),
\end{gather}
 in~\eqref{Ve}, with $\cE(x)$ defined by~\eqref{cE}, \eqref{an}, \eqref{Ed} and \eqref{cn}. This yields $V_{\rm e}(x)$ in an explicit form, in which the variables $C$, $\overline{C}$, and $\overline{\lambda}$ are no longer present. Correspondingly, rewritten in the multiplicative notation, we obtain the following
expression for $R(z)=\sum\limits_{i=1}^3S_i(z)$ where the function $\overline{\cF}(z)$ is eliminated
\begin{gather*}
\frac{R(z)}{\cF(z) \cF(q z)}=
\frac{ \big[\frac{k}{z^2}\big]\cG(q z) U\big(\frac{k}{q z}\big)}{\cG\big(\frac{k}{q z}\big) \cF(q z)}
+ \frac{\big[\frac{k}{q z^2}\big]\cG\big(\frac{k}{z}\big)U(z)}{\cG(z)\cF(z)}
+\sum_{i=1}^2\frac{k \big[\frac{k}{\ell}\big]\big[\frac{k}{z^2}\big] \big[\frac{k}{q^2 z^2}\big] \big[\frac{k}{qz^2}\big]U(\xi _i)}{\xi _i^2 \big[\frac{\xi _i^2}{\ell}\big] \big[\frac{\xi _i}{z}\big] \big[\frac{k}{\xi _i q z}\big]\cF(\xi _i)}.
\end{gather*}

We continue by adding some comments concerning genericity and questions of existen\-ce/uni\-que\-ness. We begin by noting that there are tacit genericity assumptions throughout the above. This is already the case for the van Diejen operator $A_+(\gamma;x)$: For a fixed $a_+$, we should require that $a_-$ be such that
\begin{gather*}
R_+({\rm i}a_-/2+{\rm i}a_+/2)\ne 0,\qquad R_+({\rm i}a_-+{\rm i}a_+/2)\ne 0,
\end{gather*}
to ensure that the poles of $V(\gamma;\pm x)$ and $V_{\rm b}(\gamma;x)$ are simple, and that our choice of the additive constant in $V_{\rm b}(\gamma;x)$ is finite, cf.~\cite{R15}. In particular, this says that $a_-$ should not be a multiple of~$a_+$ (equivalently, $q\ne p^n$ with $n\in\N^{*}$).

Let us next reconsider the evolution of $C$ and $\lambda$ with regard to genericity. Obviously we need to require that $C$ and $\lambda$ be nonzero, just as the variables~$\xi_n$ and the parameters $\ell$, $k$, $q$, $p$, $a_j$ and $b_j$. The above explicit solution to~\eqref{evreq} then implies that $\overline{C}$ and $\overline{\lambda}$ are nonzero as well, but this solution is only well defined when we require
\begin{gather*}
{\rm i}\phi_n\pm {\rm i}\gamma_8 \ne 0,\qquad n=1,2,\quad \mathrm{mod}\ \Lambda,
\end{gather*}
so that $c_1$ and $c_2$ are finite, cf.~\eqref{cn},
\begin{gather*}
{\rm i}\phi_n+{\rm i}a_-/2-{\rm i}a_+/2 \ne 0,\qquad n=1,2, \quad \mathrm{mod}\ \Lambda,
\end{gather*}
cf.~\eqref{ev2}--\eqref{an}, and
\begin{gather}\label{phicon}
{\rm i}(\phi_1-\phi_2)\ne 0, \qquad{\rm i}(\phi_1+\phi_2+a_-)\ne 0,\quad \mathrm{mod}\ \Lambda,
\end{gather}
cf.~\eqref{cE}.

Provided these requirements are satisfied, it is likely that the solution is also unique. Indeed, assuming there are two solutions $\cE_1(x)\ne \cE_2(x)$ of the above form, the difference would be an even degree-two elliptic function with a double pole at the origin, which vanishes for $x=i\alpha_1$ and $x={\rm i}\alpha_2$. By~\eqref{phicon} this implies that the zero sum given by (cf.~\eqref{an})
\begin{gather*}
{\rm i}(\phi_1+\phi_2 +a_--a_+),
\end{gather*}
is not congruent to 0 mod $\Lambda$, in contradiction to ellipticity. (To be sure, it is conceivable that there exists a second solution with a different structure.)

Finally, let us come back to $Z(x)$, assuming from now on $\gamma_8=\gamma_7$.
As we have shown above, in that case $Z(x)$ amounts to $V_{\rm b}(\tilde{\gamma};x)-E$, where $E$ is an unspecified constant depending on $\phi_1$. We now detail the three summands of
\begin{gather*}
Z(x)= E(x)+E(-x)+V_{\rm e}(x),\qquad \gamma_8=\gamma_7.
\end{gather*}
 From \eqref{E} we get
\begin{gather*}
E(x)= - \exp(-8{\rm i}rx-4ra_-) \frac{R_+(x-{\rm i}\gamma_7+{\rm i}a_-/2)\prod\limits_{\mu=0}^6 R_+(x-{\rm i}\gamma_{\mu}-{\rm i}a_-/2)}{R_+(2x+{\rm i}a_+/2)R_+(2x+{\rm i}a_+/2-{\rm i}a_-)}\nonumber\\
\hphantom{E(x)=}{} \times \prod_{j=1}^2\frac{R_+(x+{\rm i}\phi_j-{\rm i}a_-/2)}{R_+(x-{\rm i}\phi_j-{\rm i}a_-/2)},
\end{gather*}
whereas \eqref{Ve} combined with \eqref{evreq}--\eqref{subst} yields
\begin{gather*}
V_{\rm e}(x)= \frac{ R_+(x\pm {\rm i}(\gamma_7-a_-/2))}{\prod\limits_{j=1}^2 R_+(x\pm {\rm i}(\phi_j+a_-/2))}
\\
\hphantom{V_{\rm e}(x)=}{} \times
\left(d_1 \frac{R_+(x\pm {\rm i}(\alpha_2+a_+/2))}{R_+({\rm i}\alpha_1\pm {\rm i}(\alpha_2+a_+/2))}
+ d_2 \frac{R_+(x\pm {\rm i}(\alpha_1+a_+/2))}{R_+({\rm i}\alpha_2\pm {\rm i}(\alpha_1+a_+/2))}
\right),
\end{gather*}
with
\begin{gather*}
 d_n\equiv \exp(8r\phi_n-4ra_+) R_+\left(\frac{{\rm i}}{2}\sum_{\mu=0}^7\gamma_{\mu}-3{\rm i}a_+/2+{\rm i}a_- \pm {\rm i}a_-/2\right)
 \nonumber\\
\hphantom{d_n\equiv}{}
 \times \frac{ \prod\limits_{\mu=0}^6R_+({\rm i}\phi_n-{\rm i}\gamma_{\mu}-{\rm i}a_+/2)}{R_+({\rm i}\phi_n+ {\rm i}\gamma_7-{\rm i}a_+/2) }.
 \end{gather*}
 Here, we also used (cf.~\eqref{phis})
 \begin{gather*}
 \phi_1+\phi_2 = -a_++\frac12 a_- +\frac12 \sum_{\mu=0}^7 \gamma_{\mu},
 \end{gather*}
 and for completeness, we repeat~\eqref{an}
 \begin{gather*}
\alpha_n= \phi_n+a_-/2-a_+/2 ,\qquad n=1,2.
\end{gather*}

It should be stressed that $Z(x)$ as just obtained gives rise to a novel and quite surprising representation for $V_{\rm b}(\tilde{\gamma};x)-E$. Taking the parameters $r$, $a_+$, $a_-$ positive from now on and~$a_-$ not equal to a multiple of $a_+$ (as in~\cite{R15}), the function $V_{\rm b}(\tilde{\gamma};x)$ is real-valued for real $\gamma_{\mu}$'s and real~$x$. Taking also $\phi_1$, $\phi_2$ real, the same is true for~$Z(x)$, as is readily verified. Hence the constant $E$ is real. We can say more about its dependence on $\phi_1$, as follows.

Assuming
\begin{gather*}
2{\rm i}\gamma_7 \notin \Lambda,
\end{gather*}
we can take $x$ equal to
\begin{gather*}
x_s := -{\rm i}\gamma_7+{\rm i}a_-/2-{\rm i}a_+/2.
\end{gather*}
The point is that this yields
\begin{gather*}
E(-x_s)=V_{\rm e}(x_s)=0,
\end{gather*}
so that
\begin{gather}
Z(x_s)= - \exp(-8r\gamma_7-4ra_+) \frac{\prod\limits_{\mu=0}^6R_+({\rm i}\gamma_7+{\rm i}\gamma_{\mu}+{\rm i}a_+/2)}{ R_+(2{\rm i}\gamma_7+{\rm i}a_+/2)}\nonumber
\\
\hphantom{Z(x_s)=}{} \times \prod_{j=1}^2\frac{R_+({\rm i}\gamma_7-{\rm i}\phi_j+{\rm i}a_+/2)}{R_+({\rm i}\gamma_7+{\rm i}\phi_j+{\rm i} a_+/2)}.\label{Zsp}
\end{gather}
Now $V_{\rm b}(\tilde{\gamma};x_s)$ is a real number not depending on $\phi_1$, so we can conclude from~\eqref{Zsp} that $E$ varies over all of $\R$ as $\phi_1$ varies over~$\R$. Indeed, for $\phi_1=-\gamma_7$ the denominator in~\eqref{Zsp} vanishes, and for $\phi_1=\gamma_7$ the numerator vanishes, so by a continuity argument our conclusion readily follows.

As another salient fact, we recall from~\cite{R15} that $V_{\rm b}(\tilde{\gamma};x)$ vanishes when we choose
\begin{gather*}
\gamma_0=0,\qquad \gamma_1={\rm i}\pi/2r,\qquad \gamma_2=a_+/2,\qquad \gamma_3=a_+/2 +{\rm i}\pi/2r.
\end{gather*}
Therefore $Z(x)$ does not depend on $x$ for this $\gamma$-choice, entailing $Z(x)=Z(x_s)=-E$.

\subsection*{Acknowledgements}

The research reported in this article was initiated during a visit of S.R.\ to the Department of Mathematics of Kobe University. He would like to thank the Department for its hospitality and financial support.
This work is supported by JSPS KAKENHI Grant Numbers 26287018 and 15H03626.

\pdfbookmark[1]{References}{ref}
\LastPageEnding

\end{document}